\documentclass{pnastwo}

\newcommand\beq{\begin{equation}}
\newcommand\eeq{\end{equation}}
\newcommand\beqa{\begin{eqnarray}}
\newcommand\eeqa{\end{eqnarray}}

\begin{document}

\title{Atypical late-time singular regimes accurately diagnosed in stagnation-point-type solutions of 3D Euler flows}

\author{Rachel M. Mulungye\affil{1}{Complex and Adaptive Systems Laboratory, School of Mathematics and Statistics, University College Dublin, Belfield, Dublin 4, Ireland},
 Dan Lucas\affil{1}{}
\and
  Miguel D. Bustamante\affil{1}{}}

\maketitle

\begin{article}
\begin{abstract}
{We revisit, both numerically and analytically, the finite-time blowup
of the infinite-energy solution of 3D Euler equations of
stagnation-point-type introduced by Gibbon et al. (1999).  By
employing the method of mapping to regular systems, presented in
Bustamante (2011) and extended to the symmetry-plane case by Mulungye
et al. (2015), we establish a curious property of this solution that
was not observed in early studies: before but near singularity time,
the blowup goes from a fast transient to a slower regime that is well
resolved spectrally, even at mid-resolutions of $512^2.$ This
late-time regime has an atypical spectrum: it is Gaussian rather than
exponential in the wavenumbers. The analyticity-strip width decays to
zero in a finite time, albeit so slowly that it remains well above the
collocation-point scale for all simulation times $t < T^* -
10^{-9000}$, where $T^*$ is the singularity time. Reaching such a
proximity to singularity time is not possible in the original temporal
variable, because floating point double precision ($\approx 10^{-16}$)
creates a `machine-epsilon' barrier. Due to this limitation on the
\emph{original} independent variable, the mapped variables now provide
an improved assessment of the relevant blowup quantities, crucially
with acceptable accuracy at an unprecedented
closeness to the singularity time: $T^*- t \approx 10^{-140}.$}

\end{abstract}


\abbreviations{3D, three-dimensional; ODE, ordinary differential equation; GPU, graphics processing unit; BKM, Beale-Kato-Majda}

\dropcap{T}he open question of regularity of the fluid dynamical equations is considered one of the most fundamental challenges of mathematics and physics \cite{fefferman:2006}. While the viscous Navier-Stokes equations have more physical relevance, the inviscid Euler equations present the greatest challenge and exhibit the most extreme behaviours. For this reason the numerical study of possible finite-time blowup is typically concerned with these inviscid equations. The three-dimensional Euler  equations for an incompressible fluid of unit mass density with velocity field $\mathbf{u}(x,y,z,t) \in \mathbb{R}^3$, in a time interval $t \in [0,T),$ can be expressed as:
\begin{eqnarray}
\label{eq:Euler}
 \frac{\partial \mathbf{u}}{\partial t} + \mathbf{u} \cdot \nabla \mathbf{u} = - \nabla p\,, \qquad \quad \nabla \cdot \mathbf{u} = 0.
\end{eqnarray}
Periodic boundary conditions are commonly assumed in a fundamental domain $[0,2\pi]^3\,.$ Beale \emph{et. al.} \cite{Beale:1984p2225} provided a blowup criterion based on the maximum vorticity (infinity norm) providing an important reference for diagnosing singularities numerically (BKM theorem). The literature surrounding the numerical assessment of finite-time blowup in 3D Euler is extensive and will not be reviewed in detail here (see \cite{Bardos:2007gf}, \cite{Gibbon:2008p1805} and for references pertinent to the current work see the introduction in \cite{Mulungye:2015fs}), suffice to say that a fundamental difficulty of this important problem is the lack of analytic solutions or any \emph{a priori} knowledge of asymptotic behaviour. A secondary obstacle is that the spatial collapse associated with intense vortex stretching results in numerical solutions becoming unresolved beyond a certain time (e.g., loss of spectral convergence). It is therefore these authors' view that it is imperative to devise a framework with nontrivial blowup dynamics and where analytic solutions are known in order to validate and compare various numerical methods, for the purposes of accurately solving the system \emph{and} diagnosing blowup.
In this regard we reinvestigate the stagnation-point-type solutions (with infinite energy) of the $3$D Euler equations \cite{Gibbon:1999uo} found to have analytic solutions exhibiting finite-time blowup \cite{Constantin:2000fa}, now employing a novel new method that maps nonlinearly the time and fields to a globally regular system \cite{Bustamante:2011da}. By doing this we uncover curious late-time behaviour of the Fourier spectrum such that in the new variables the solution remains well spectrally converged until far beyond the time at which the original system could hope to achieve due to floating point precision proximity to the singularity time $T^*.$ We find that the mapped variables maintain acceptable levels of error in the main blowup quantities such as the $L^\infty$ and $L^2$ norms of the vortex stretching rate at this extreme closeness to $T^*.$ We begin by formulating both the original and mapped equations and reviewing and updating the analytic and asymptotic results for blowup. We then present a thorough investigation of the Fourier spectra of the solution, followed by error analysis of our numerics and an assessment of singularity time and proximity to it. 


\section{Formulation}
We consider a class of exact solutions of the 3D Euler equations presented by Gibbon et al.~\cite{Gibbon:1999uo}. Writing $\mathbf{u}(x,y,z,t) = (u_x(x,y,t), u_y(x,y,t), z\,\gamma(x,y,t))$ we obtain
\begin{eqnarray}
\label{GAMMA2}
\frac{\partial \gamma}{\partial t} +\mathbf{u}_\mathrm{h}\cdot\nabla_\mathrm{h} \gamma &=& 2\langle\gamma^2\rangle - \gamma^2\,,\\
\label{OMEGA3}
\frac{\partial \omega}{\partial t} +\mathbf{u}_\mathrm{h}\cdot\nabla_\mathrm{h} \omega &=& \gamma\,\omega\,,
\end{eqnarray}
where $\mathbf{u}_\mathrm{h}(x,y,t) \equiv (u_x(x,y,t), u_y(x,y,t))$ denotes the ``horizontal'' component of the velocity field at the symmetry plane ($z=0$), $\nabla_\mathrm{h} = (\partial_x, \partial_y)$ denotes the ``horizontal'' gradient operator, $\omega$ is the vorticity defined as
$$\omega(x,y,t) = \partial_x u_y - \partial_y u_x,$$
$\gamma$ is the stretching-rate of vorticity, which using the incompressibility condition in equation~(\ref{eq:Euler}) can be defined as: 
\begin{eqnarray}
\label{eq:def_gamma}
\gamma(x,y,t) = - \nabla_\mathrm{h} \cdot \mathbf{u}_\mathrm{h}(x,y,t)\,,
\end{eqnarray}
and 
$$\langle f(x,y,t) \rangle \equiv \frac{1}{4\pi^2} \int_0^{2\pi}\int_0^{2\pi}f(x,y,t)~dx~dy$$
denotes the spatial average over the periodic $2$D domain.
 
Constantin \cite{Constantin:2000fa} solved for $\gamma$ along characteristics (and for vorticity $\omega$, which can be found a posteriori), proving that the stretching rate $\gamma$ would blow up in a finite time, with explicit formulae for the singularity time which confirmed the accuracy of the numerical blowup predictions in \cite{Ohkitani:2000wq}. A BKM \cite{Beale:1984p2225} type of theorem was established by Gibbon \cite{Gibbon:2001bc} where the blowup time $T^*$ is defined as the smallest time at which 
$$\int_0^{T^*} \|\gamma(\cdot,t')\|_\infty~dt' = \infty,$$ 
where $\|\gamma(\cdot,t)\|_\infty$ is the supremum norm of the vorticity stretching rate. 
 
Bustamante \cite{Bustamante:2011da} and later Mulungye \textit{et. al.} \cite{Mulungye:2015fs} introduced the following `mapped' fields and `mapped' time: 
\begin{eqnarray}
\nonumber 
\gamma_{\mathrm{map}}(x,y,\tau) &=& \frac{\gamma(x,y,t)}{\|\gamma(\cdot,t)\|_\infty}, \qquad \tau(t) = \int_0^t \|\gamma(\cdot,t')\|_\infty~dt',\\
\label{eq:mapping}
\omega_{\mathrm{map}}(x,y,\tau) &=& \frac{\omega(x,y,t)}{\|\gamma(\cdot,t)\|_\infty}, 
\end{eqnarray}
This transformation is bijective for $t<T^*.$ The mapped fields satisfy the following PDE system:
\begin{eqnarray}
\label{PDE1}
\frac{\partial\gamma_{\mathrm{map}}}{\partial\tau}+\mathbf{u}_{\mathrm{map}}\cdot\nabla \gamma_{\mathrm{map}}
\hspace{-3mm}&=&\hspace{-3mm} 2 \langle\gamma_{\mathrm{map}}^2\rangle - \gamma_{\mathrm{map}}^2\nonumber\\
&+&\hspace{-4mm}\sigma_\infty\gamma_{\mathrm{map}}\hspace{-1mm}\left\{1
\hspace{-0.5mm}- 2 \langle\gamma_{\mathrm{map}}^2\rangle\hspace{-0.5mm}\right\}\\
\label{PDE2}
\frac{\partial\omega_{\mathrm{map}}}{\partial\tau} +\mathbf{u}_{\mathrm{map}}\cdot\nabla \omega_{\mathrm{map}}
\hspace{-3mm}&=&\hspace{-3mm}\gamma_{\mathrm{map}}\,\omega_{\mathrm{map}}\nonumber\\
&+&\hspace{-4mm}\sigma_\infty\omega_{\mathrm{map}}\hspace{-1mm}\left\{1 
\hspace{-0.5mm}- 2 \langle\gamma_{\mathrm{map}}^2\rangle\hspace{-0.5mm}\right\}
\end{eqnarray}
where $\sigma_\infty \equiv \mathrm{sign} \, \gamma(\mathbf{X}_{\gamma}(t),t)
$ is the sign of $\gamma$ at the position $\mathbf{X}_{\gamma}(t)$ of maximum $|\gamma(\mathbf{x},t)|$.
The initial conditions used in this study are:
\begin{equation}
\label{eq:initialC}
\gamma_0(x,y)= \omega_0(x,y)= \sin(x)\sin(y)\,.
\end{equation}


\subsection{Analytical solution of the stagnation-point-type 3D Euler flows}
\label{sec:SP_ana}

System (\ref{GAMMA2})--(\ref{OMEGA3}) is an exact solution of $3$D Euler equations (albeit with infinite energy), as derived originally by Gibbon \textit{et. al.} \cite{Gibbon:1999uo}.  Ohkitani and Gibbon \cite{Ohkitani:2000wq} performed a numerical study at resolution $256^2,$ supported with simulations at resolution $1024^2,$ which provided evidence of a finite-time singularity at $t \approx 1.4.$ Higher resolution was not needed due to the fact that spectral convergence was observed during most of the simulation time. 

Constantin \cite{Constantin:2000fa} introduced a method for finding analytically the blowup quantities (e.g. $\|\gamma(\cdot,t)\|_\infty, \, \langle \gamma^2\rangle$) and established that there is a finite-time singularity. While it is possible to obtain the asymptotic behaviour of the blowup quantities using the notation in \cite{Constantin:2000fa}, we will discuss this in the context of our method \cite{Mulungye:2015fs} for the purposes of simplicity of presentation. The equation of motion for $S(t),$ the function upon which provides the solution for all fields along characteristics, (see \cite{Mulungye:2015fs} for more details) is, following spatial integration over the initial condition:
\beq\dot{S} = \frac{\pi^2}{4\,\left[K(S^2)\right]^2}\,, \quad S(0) = 0,\label{eq:Seq} \eeq
where $K(\mu^2)$ is the complete elliptic function of the first kind such that $S$ can be interpreted in this case as the modulus of the elliptic function. The singularity occurs when $S = S^*$, where $S^* \equiv  -1/\inf\gamma_0 = 1.$ Thus, $S(t)$ goes from $0$ at $t=0$ to $1$ at $t=T^*$, where $T^*$ is the singularity time defined as

\begin{eqnarray}
T^* \hspace{-2mm}&=&\hspace{-2mm}\frac{4}{\pi^2} \int_0^1 \left[K(S^2)\right]^2~dS \approx 1.418002734923858875062234\nonumber.
\end{eqnarray}

The singularity is dominated by the infimum of $\gamma(\mathbf{x},t)$ and for these initial conditions we can identify $\|\gamma\|_\infty = - \inf \gamma,$ so $\sigma_\infty = -1$ will be used throughout.\\

Solving the above ODE for $S(t)$ is in principle feasible numerically to any desired accuracy, although we do not know if a closed form is available. However, it is possible to obtain some exact formulae for $\|\gamma(\cdot,t)\|_\infty$ and $\tau(t)$ in implicit form through $S(t)$:

\begin{equation}
\label{eq:tau_of_S_lambda=0}
\tau(t) = - \ln\left(\frac{2}{\pi}(1-S)K(S^2)\right),
\end{equation}

\begin{equation}
\label{eq:gamma_of_S_lambda=0}
 - \inf_{\mathbf{x}\in \mathbb{T}^2} \gamma(\mathbf{x},t) = \|\gamma(\cdot,t)\|_\infty = \frac{\pi ^2 \left[(S+1) K\left(S^2\right) - E\left(S^2\right)\right]}{4 S \left(1-S^2\right) K\left(S^2\right)^3}\,,
\end{equation}


\beq\left\langle \gamma ^2\right\rangle = \frac{\pi ^4 \left[K(S^2)-E(S^2)\right]\left[2E(S^2)-\left(1-S^2\right)K(S^2)\right]}{32 S^2  \left(1 - S^2\right)^2 K\left(S^2\right)^6}\,,\eeq

Notice that it is possible to find $S(\tau)$ and $\|\gamma(\cdot,t(\tau))\|_\infty$ as functions of $\tau$ via first inverting equation (\ref{eq:tau_of_S_lambda=0}) to obtain $S(\tau),$ and then using this to obtain $\|\gamma(\cdot,t(\tau))\|_\infty.$ In numerical implementations at values of $\tau$ greater than about $33,$ this requires the use of arbitrary-precision computations, provided by commercial packages such as \emph{Mathematica} as $S(\tau)$ becomes to within double floating point precision to $1.$ 

Defining $Z \equiv - \ln\left(\frac{1-S}{8}\right),$ the following formulae are valid asymptotically as $S \lessapprox 1:$
\begin{eqnarray}
\label{asym:Tstar} T^*-t &\approx& \frac{8\mathrm{e}^{-Z}}{\pi^2}\left(Z^2 + 2 Z + 2\right)\,,\\
\label{asym:inf}  \|\gamma(\cdot,t)\|_{\infty} &\approx& \frac{\pi^2 \mathrm{e}^{Z}}{8}\left(\frac{Z-1}{Z^3}\right)\,,\\
\label{asym:gsq}  \left\langle \gamma ^2\right\rangle  &\approx& \frac{\pi^4 \mathrm{e}^{2Z}}{128}  \left(\frac{Z-2}{Z^6} \right)\, .
\end{eqnarray}
Also, we obtain $\tau \approx Z - \ln \left(\frac{8 Z}{\pi}\right)$ to lowest order. This latter formula can be inverted in the asymptotic region of interest, giving
$$Z \approx -W_{-1}\left(-\frac{1}{8} \pi  e^{-\tau }\right)\,,$$
where $W_{-1}$ is a branch of the Lambert (or Product Log) function. By combining the above it is possible to obtain explicit asymptotic expressions for $T^*-t$ and $\|\gamma(\cdot,t)\|_{\infty}$ in terms of the mapped time $\tau.$

These asymptotic formulae are very useful in practice. At $\tau = 5$ the above asymptotic formula for $t(\tau)$ has a relative error of about $10^{-9}$  and for $\|\gamma(\cdot,t)\|_{\infty},$ a relative error of $10^{-7}.$ By $\tau>20$ the asymptotic formulae above are accurate to double precision ($10^{-16}$).

\section{Numerical solution of original and mapped systems}
\label{sec:comparison}

We solve the evolution equations for both systems numerically using a standard pseudospectral method implemented on GPUs using CUDA  \cite{Mulungye:2015fs}. Dealiasing is carried out using Hou's exponential filter {$\exp\left(-36 \left(2k/N\right)^{36}\right)$ (for a given spatial resolution $N$) \cite{Hou:2006p2135} } and a fourth-order Runge-Kutta scheme solves in time. Adaptive time-stepping, ($dt = d\tau/\|\gamma(\cdot,t)\|_{\infty}$), is used for the original equations and uniform steps of $d\tau$ are used in the mapped system with the resulting distribution of temporal data roughly equivalent.

\subsection{Spectra and analyticity strip}
\label{sec:spectra}

To investigate the spatial collapse associated with the singularity in $\gamma$, we consider a detailed analysis of the one-dimensional spectra of stretching rate $\gamma$ constructed from spherical shells:
$$E(k,t)=\sum_{k-\frac{1}{2}<\left|\textbf{k}\right|< k+\frac{1}{2}}|\hat{\gamma}(\textbf{k},t)|^2.$$
Our first observation of the evolution of the spectrum is that there are two timescales in evidence. An initial burst can be observed with a flux towards intermediate $k$ which is redistributed across the modes. Provided $N>256$ this initial phase remains well resolved and lasts only until $\tau\approx 25$. Thereafter there is a {slow 
cascade} from small $k.$ In fact in original variables the initial phase is until $T^*-t \approx 10^{-10}.$ As will be shown, this is too early to establish certain asymptotic trends.

\begin{figure*}
\scalebox{0.8}{\input{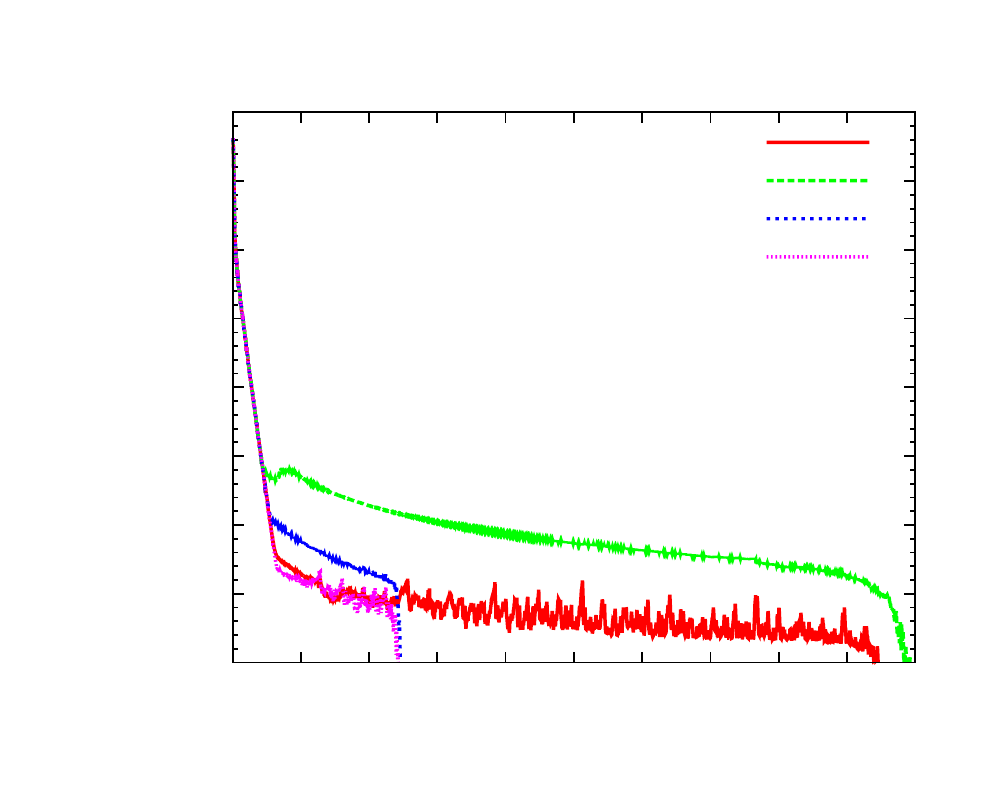}}
\scalebox{0.8}{\input{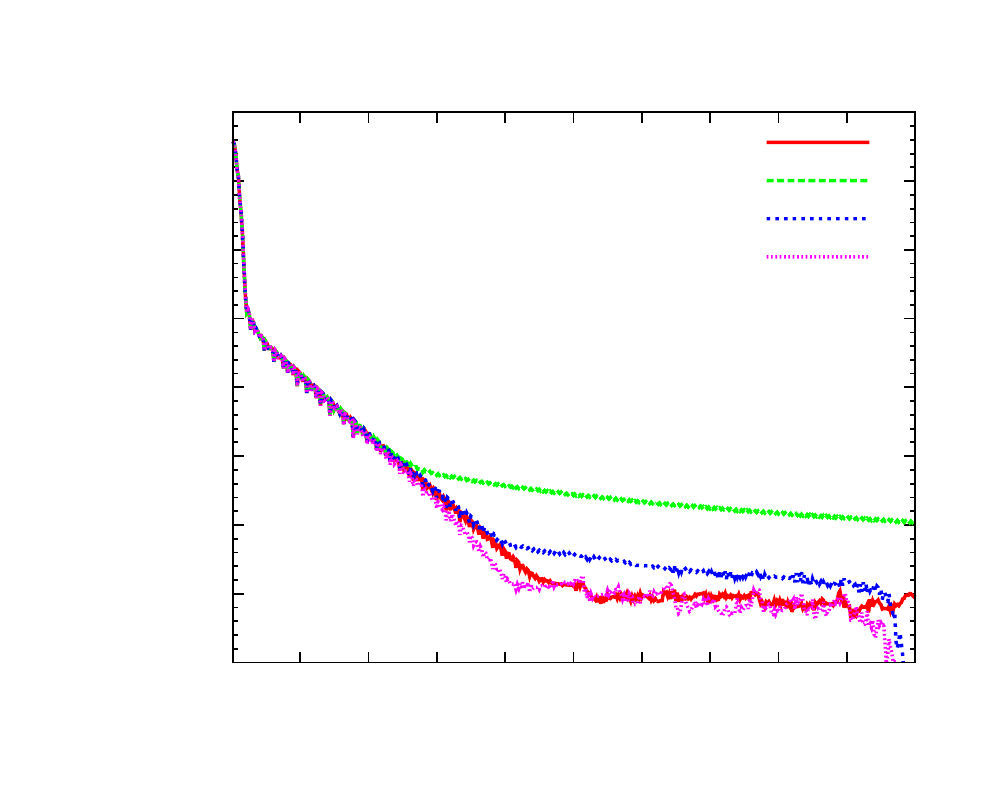}}\\

\vspace{-10mm} 
\scalebox{0.8}{\input{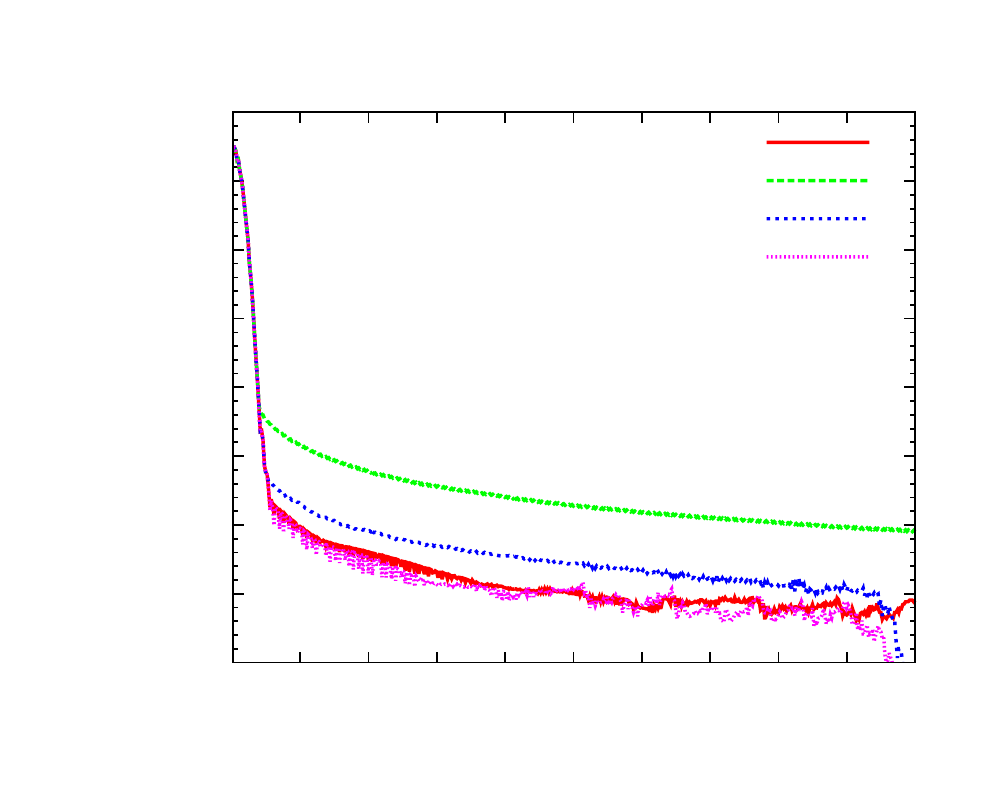}}
\scalebox{0.8}{\input{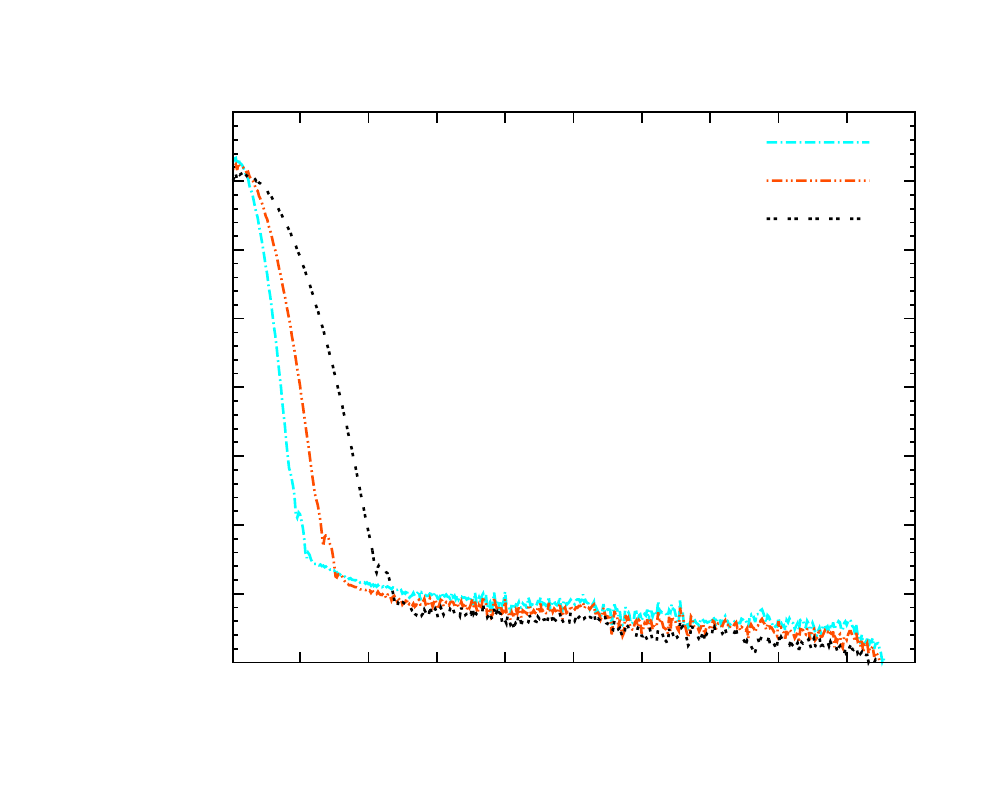}}
\caption{Snapshots of spectra for $\tau=5,10$ and $25$ (top left and right and bottom left respectively), on a lin-log scale. These first three figures show two resolutions ($N=1024$ and $4096$), with and without hyperviscosity, demonstrating the need to control floating point round-off error at small scales. $\tau=5$ shows the full spectrum, including dealiased filtered modes to show the small scale error. Thereafter, $\tau=10$ and $25$ plots show only the first 500 modes to make clear the initial burst and the onset of the slow Gaussian spectrum. The final frame (bottom right) shows only the $N=1024$ case with hyperviscosity, now with curves at $\tau=100,$ $200$ and $500$ showing the slow broadening of the spectrum at late times.\label{fig:spec}}
\end{figure*}

\begin{figure}
\centering
\scalebox{0.6}{\input{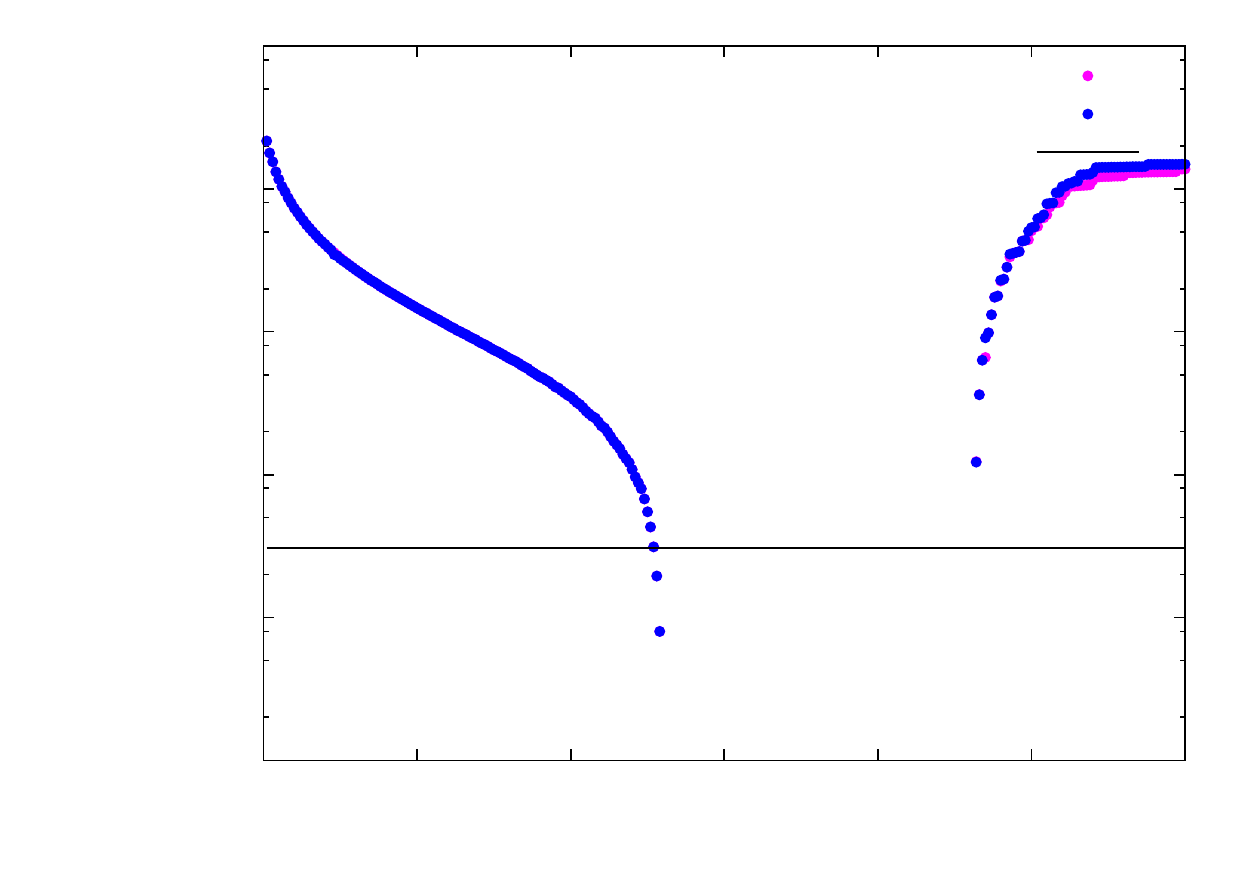}}
 \scalebox{0.6}{\input{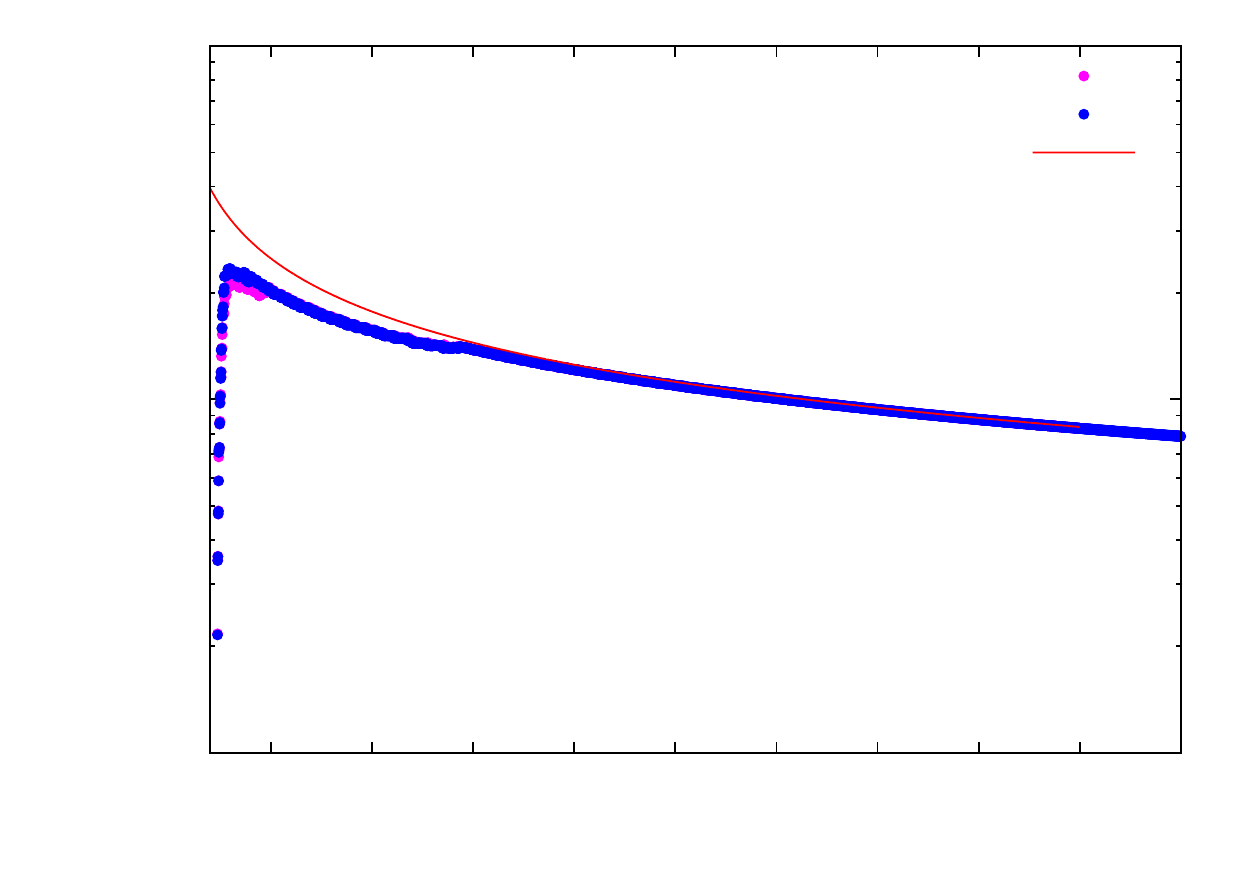}}
\caption{ Time evolution of the analyticity distance $\delta (\tau)$ of the Fourier spectrum of $\gamma^2$. Top: profile at early times based on equation (\ref{eq:fit1}), showing the initial burst. Bottom: late time profile based on equation (\ref{eq:fit2}) showing the slow Gaussian cascade along with the estimate $\delta \approx \sqrt{\pi/ \tau}$, which saturates inequality (\ref{ieq:delt}). \label{fig:delt}}
\end{figure}

It was also found that, due to the lack of direct energy cascade to large $k$, an accumulation of round-off error propagates up-scale. The result is a small quantity of spurious energy between the large scales and the truncation wavenumber. The amount of this spurious energy is resolution dependent leading to an ill-converged spectrum.  This issue was remedied by applying a small amount of hyperviscosity on the large wavenumbers, namely adding the term 
\begin{equation}
\nu (-1)^{2h+1}|\mathbf{k}|^{2h} \hat{\gamma}, \qquad  h = 2,  \qquad \mathrm{for} \, |\mathbf{k}|>200
\end{equation}
to the right hand side of the Fourier transform of equation (\ref{PDE1}) and (\ref{PDE2}). Numerically a Crank-Nicolson scheme was used on this term for stability. Figure \ref{fig:spec} shows the profile of the spectra at $\tau=5,$ $10$ and $25$ for $N=1024$ and $4096,$ each with $\nu=10^{-9}$ and $\nu=0$. This demonstrates that the hyperviscosity gives a well converged spectrum while leaving the large scale modes unaffected. The error in the bulk quantity $\langle \gamma^2_{\mathrm{map}}\rangle$ is unchanged (figure not shown), however applying hyperviscosity to all modes leads to a significant  error increase.
 Interestingly the late time profile does not have the typical shape we might expect \cite{Bustamante:2012jc,Mulungye:2015fs} or that which is assumed previously in this system \cite{Ohkitani:2000wq}, namely 
\begin{equation}
E(k,t) = C(t)k^{-n(t)}\text{e}^{-2\delta_1(t)k}.
\label{eq:fit1}
\end{equation}
In fact, as can be seen in figure \ref{fig:spec} (bottom right at late times), the profile assumes a more Gaussian shape, 
\begin{equation}
E(k,t) = C(t)k^{-n(t)}\text{e}^{-(\delta_2(t)k)^2}.
\label{eq:fit2}
\end{equation}
This late time spatial form has been missed in previous work \cite{Ohkitani:2000wq} on this system as it only arises after the initial burst, which does have the $\text{e}^{-2\delta k}$ shape, and persists to sufficiently close to $T^*$ to render it next to inaccessible without the mapped variables. To ensure the convergence of the initial burst phase we first perform a least-squares fitting procedure to the spectrum with ansatz (\ref{eq:fit1}). Figure \ref{fig:delt} shows the fitted $\delta_1$ for the {early 
burst} phase. The plot shows two resolutions ($N=1024$ and 2048) which are essentially indistinguishable.  From Figure \ref{fig:spec} the exponential part of the profile of $E(k,t)$ is preceded (in $k$) by the Gaussian shape. The downscale flux associated with the slackening of the exponential part ($\delta_1$ decreasing) establishes the Gaussian profile in its wake. The result is that, while the trend in $\delta_1$ suggests an exponential decay in $\tau$ (at early times), in reality the ansatz (\ref{eq:fit1}) ceases to be a valid analyticity measure due to the addition of a large scale Gaussian spectrum. The cross over regime is indicated by negative values of $\delta_1$ for $13 \lesssim \tau \lesssim 23.$

To analyse the true late time behaviour ($\tau>25$) we fit with ansatz (\ref{eq:fit2}). Figure \ref{fig:delt} shows the behaviour of $\delta_2$ as a function of $\tau$. Strikingly, the decay is now very slow (see Eq.~(\ref{ieq:delt})).

\begin{figure}
\centering
\scalebox{0.6}{\input{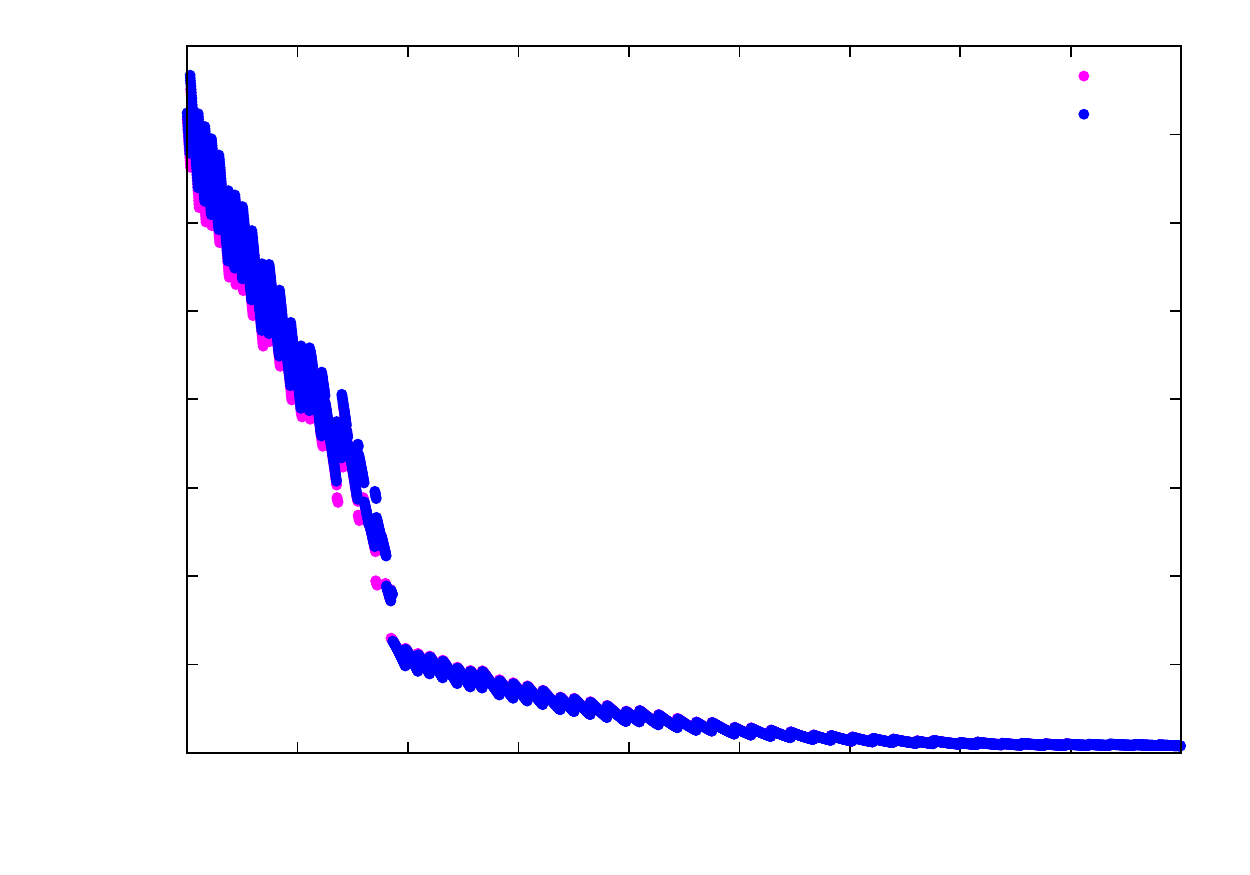}}
\scalebox{0.6}{\input{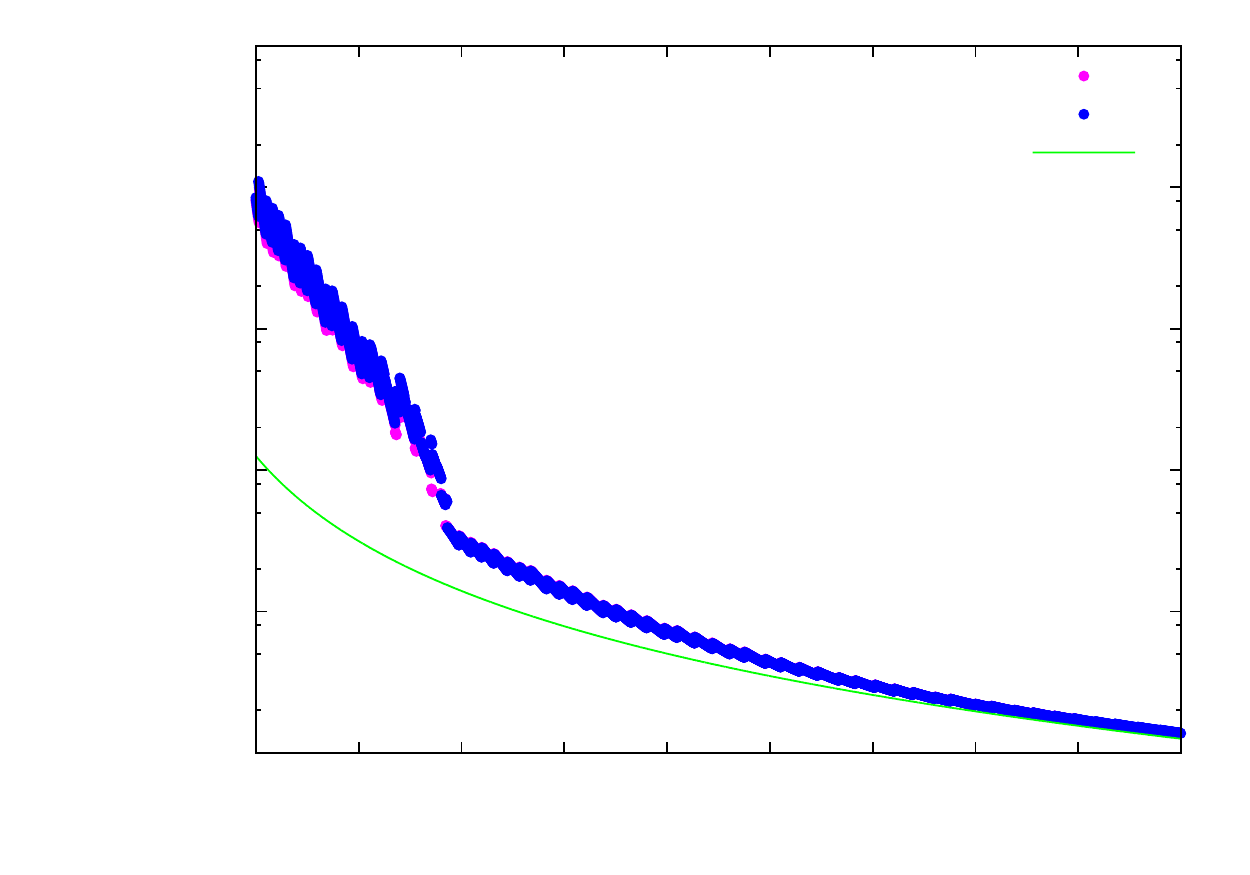}}
\caption{ Time evolution of the exponent, $n (t(\tau))$, and constant factor, $C(t(\tau))$ (normalised to coincide with the mapped variables), of the fit ansatz equation (\ref{eq:fit2}) of the Fourier spectrum of $\gamma^2$. Along side $C(t(\tau))$ is the saturated estimate $\pi \tau^{-2}$ from inequalities (\ref{eq:ineq}) and (\ref{ieq:delt}). \label{fig:nC}}
\end{figure}


Using a classical method, it is possible to obtain a rigorous upper bound for the supremum norm of stretching rate in terms of the spectrum:  
\begin{eqnarray}
\label{eq:ineq}
\| \gamma (\cdot,t)\|_\infty &\leq& \sum_{k=1}^\infty \sqrt{S_k} \sqrt{E(k,t)}\,\,,\\
\nonumber
S_k &\equiv& \# \{{\mathbf{k}} \in \mathbb{Z}_{\mathrm{odd}}^2 \cup \mathbb{Z}_{\mathrm{even}}^2:\,  k-1/2  < |{\mathbf{k}}| < k+1/2 \} \\
\nonumber
 & \approx& \pi \, k ,\qquad k \to \infty.
\end{eqnarray}

The special condition on odd-odd or even-even modes is due to the discrete symmetry of our initial condition.

Replacing the fit (\ref{eq:fit2}) into (\ref{eq:ineq}) leads to a bound involving an infinite sum over $k$ with an ultraviolet divergence in the limit of small $\delta(t).$ We can approximate this as follows \cite{Bustamante:2012jc}:
\begin{equation*}
\| \gamma (\cdot,t)\|_\infty \leq \frac{1}{2} \sqrt{\pi C(t)}\, \Gamma\left(1-\frac{n(t)+1}{4}\right) \,\left[\frac{1}{2}\delta(t)^2\right]^{\frac{n(t)+1}{4}-1},
\end{equation*}
where $\Gamma$ is the gamma (factorial) function. A further improvement is obtained by noticing the behaviour of $n(t)$ at late times from figure \ref{fig:nC}, where it is clear that $n(t) \to -1.$ Therefore we obtain, in this limit,
\begin{equation}
\label{eq:ineq_2}
\| \gamma (\cdot,t)\|_\infty \leq \sqrt{\pi C(t)} \, \delta(t)^{-2}.
\end{equation}
This inequality alone cannot be used to estimate the behaviour of $\delta(t),$ since the independent factor $C(t)$ is involved as well, so an extra equation is needed. This extra equation is provided by combining the asymptotic formulae (\ref{asym:inf}) and (\ref{asym:gsq}):
$$\left\langle \gamma(\cdot,t)^2\right\rangle  \approx \frac{\| \gamma (\cdot,t)\|_\infty^2}{2\, \tau}.$$
The left-hand-side of this equation can be written in terms of the energy spectrum, so if we follow similar steps as in the derivation of inequality (\ref{eq:ineq_2}) we obtain $\left\langle \gamma(\cdot,t)^2\right\rangle  \approx \frac{C(t)}{2} \delta(t)^{-2}.$ Therefore we get
\begin{equation}
\label{eq:C(t)}
C(t) \approx \frac{\| \gamma (\cdot,t)\|_\infty^2\,\delta(t)^{2}}{\tau}.
\end{equation}
Using this we can go back to inequality (\ref{eq:ineq_2}) and show that it is equivalent to:
\beq \label{ieq:delt} \delta(t) \leq \sqrt{\frac{\pi}{\tau}}.\eeq
This inequality is in fact saturated, as confirmed by our numerical simulation (figure \ref{fig:delt}). In terms of original time variable we get
$$\delta(t) \leq \sqrt{\frac{\pi}{-W_{-1}(-\pi(T^*-t))}} \approx  \sqrt{\frac{\pi}{-\ln(\pi(T^*-t))}},$$
 which illustrates that the loss of regularity is very slow in the original time variable. In fact if one were to consider the reliability time with the saturated spectrum one would find that $\tau_{rel} \approx \frac{N^2}{4\pi},$ so that for $N=256,$ $\tau_{rel}\approx 2\times10^5$ or $T^*-t \approx 10^{-9000}.$
\subsection{Errors} 
Using the definition in \cite{Mulungye:2015fs} the `normalised' $L^2$ norm of the error (\emph{not} relative error) is given by
 \[Q(f,g) = \frac{\| f - g \|_2}{\|f\|_2+\|g\|_2}. \]
We consider the error associated with the local quantity $\|\gamma(\cdot,t)\|_{\infty}$ and the global ones $\langle\gamma_{\mathrm{map}}^2\rangle, \langle\gamma^2\rangle$ via $Q_\gamma = Q(\|\gamma_{\mathrm{num}}(\cdot,t)\|_{\infty},\|\gamma_{\mathrm{ana}}(\cdot,t)\|_{\infty})$ and $Q_{\langle\gamma^2\rangle} = Q(\langle\gamma^2_{\mathrm{num}}\rangle,\langle\gamma^2_{\mathrm{ana}}\rangle),$ etc. where the subscripts ``num'' and ``ana'' stand for ``numerical'' and ``analytic''.

\begin{figure}
\centering
\scalebox{0.55}{\input{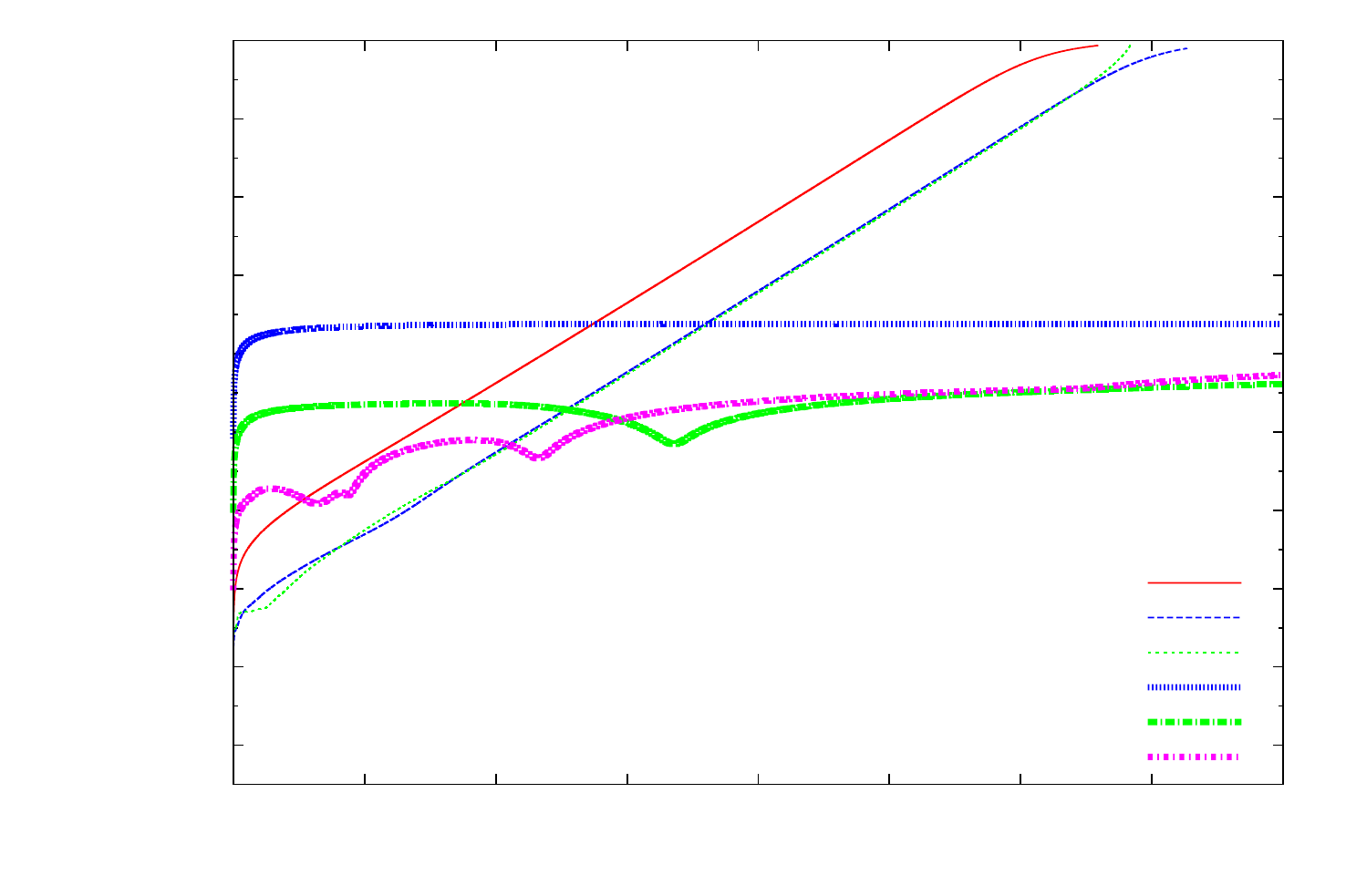}}

\vspace{-8mm}
\scalebox{0.55}{\input{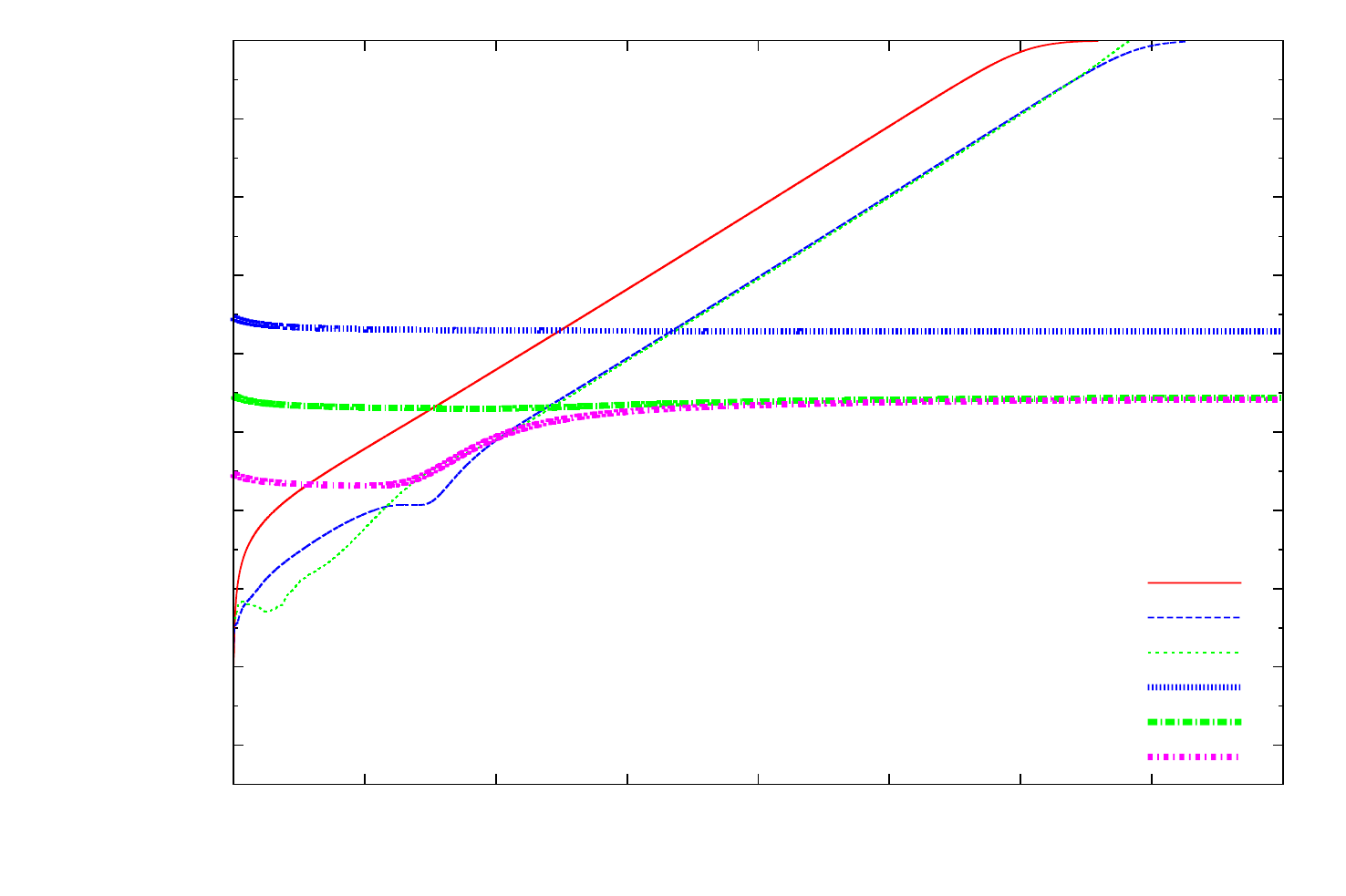}}

\caption{Time evolution of the error measures $Q_\gamma$, (top) $Q_{\langle\gamma_{\mathrm{map}}^2\rangle}$ and $Q_{\langle\gamma^2\rangle}$ (bottom) showing the convergence with time step $d\tau.$ \label{fig:err}}
\end{figure}

The numerical solution of the mapped system does not provide direct access to the original variable $\|\gamma(\cdot,t)\|_\infty$ so the following expression is required \cite{Mulungye:2015fs}
\beq
\label{eq:norm_gamma}
\|\gamma(\cdot,t(\tau))\|_\infty =\|\gamma_0\|_\infty \exp\left[\tau -
2 \int_0^{\tau} \langle\gamma_{\mathrm{map}}^2\rangle \mathrm{d}\tau'\right]
\eeq
 
\noindent where $\int_0^{\tau}\langle\gamma_{\mathrm{map}}^2\rangle \mathrm{d}\tau'$ is computed using Simpson's rule. We compare both this mapped estimate and the direct supremum norm from the original system against the analytical solution of Eq.~(\ref{eq:gamma_of_S_lambda=0}). Care is taken in solving Eq.~(\ref{eq:Seq}) so that the time steps from the original system are used to solve on intervals which coincide with the data points and arbitrary precision of the required level is used. As shown in the previous section the solution remains well resolved spatially, even at relatively modest resolutions, therefore we omit the error study of spatial convergence here. We do however show convergence with respect to timestep $d\tau$ in figure \ref{fig:err} at resolution $N=1024$.  Overall we observe an exponential growth (in $\tau$) of $Q_\gamma$ from the original system, compared to an almost uniform error ($\sim 10^{-10}$) from the mapped version at converged $d\tau$. Interestingly convergence occurs in the mapped system at a smaller level of $d\tau.$

In contrast to the earlier result in \cite{Mulungye:2015fs}, we find that both  $Q_\gamma$ and $Q_{\langle\gamma_{\mathrm{map}}^2\rangle}$ behave similarly; recall that $\langle\gamma_{\mathrm{map}}^2\rangle$ is the primary variable for the assessment of the local quantity $\|\gamma(\cdot,t)\|_{\infty}$ in the mapped system (equation (\ref{eq:norm_gamma})). This implies that in this case (where $\langle \gamma^2\rangle$ is not an invariant), $Q_\gamma$ is simply slaved to $Q_{\langle\gamma_{\mathrm{map}}^2\rangle}.$ Ref.~\cite{Mulungye:2015fs} contains a detailed discussion on error sources in the mapped and original variables, and highlights some subtleties surrounding the behaviour of $\langle \gamma^2\rangle.$ Here the situation is somewhat more straightforward: the original system contains unbounded error growth due to a fundamental loss of precision in the independent variable. This is explained by the earlier convergence and the `saturation' of error near the double-precision limit ($\tau\approx37$). 

\subsection{Assessing blowup time $T^*$ and proximity to it, $T^*-t$}
Previous methods for assessing the value of $T^*,$ e.g. fitting the behaviour of $\|\gamma(\cdot,t)\|_\infty $ to a power law $(T^*-t)^\alpha,$ are based on the assumption that the solution is incurring significant errors at intermediate time and $T^*$ requires careful extrapolation. Here the solution is remaining well resolved until late times and we find that using the original system with the adaptive timestep given above, $t$ converges to $T^*$ to within $\sim10^{-14}.$ This accuracy is surprising given it arises from a simple sum $t_i = \sum_i d\tau/\|\gamma(\cdot,t_i)\|_\infty,$ and it cannot be improved by fitting or even by arbitrary precision arithmetic to sum $dt$ which are below the machine precision threshold. Understanding this accuracy is aided by attempting the comparable exercise for the mapped system. Here the recovery of $t$ is given by
\[ t(\tau) = \int_0^\tau \frac{1}{\|\gamma(\cdot,t(\tau'))\|_\infty} \mathrm{d} \tau'\,,\]
{where $\|\gamma(\cdot,t(\tau'))\|_\infty$ is obtained from formula (\ref{eq:norm_gamma}). This integral should converge to $T^*$} as $\tau\to \infty$. Numerically computing it results in a saturation of error $\sim10^{-9}$ {(slaved to the error in $\|\gamma(\cdot,\tau)\|_\infty$)} when $d\tau=10^{-4}$ for sufficiently large $\tau$. This almost leads to a paradox: why should a quantity with lower late time error produce a poorer estimate for $T^*$ when the procedure for the estimate is qualitatively the same. The reason is that it is the \emph{early} errors which pollute the estimate for $t(\tau)$ as these are larger in the mapped system and occur at a point where they will contribute more significantly to the final integral ($d\tau/\|\gamma(\cdot,t)\|_\infty $ is largest). 

\begin{figure}
\centering
\scalebox{0.6}{\input{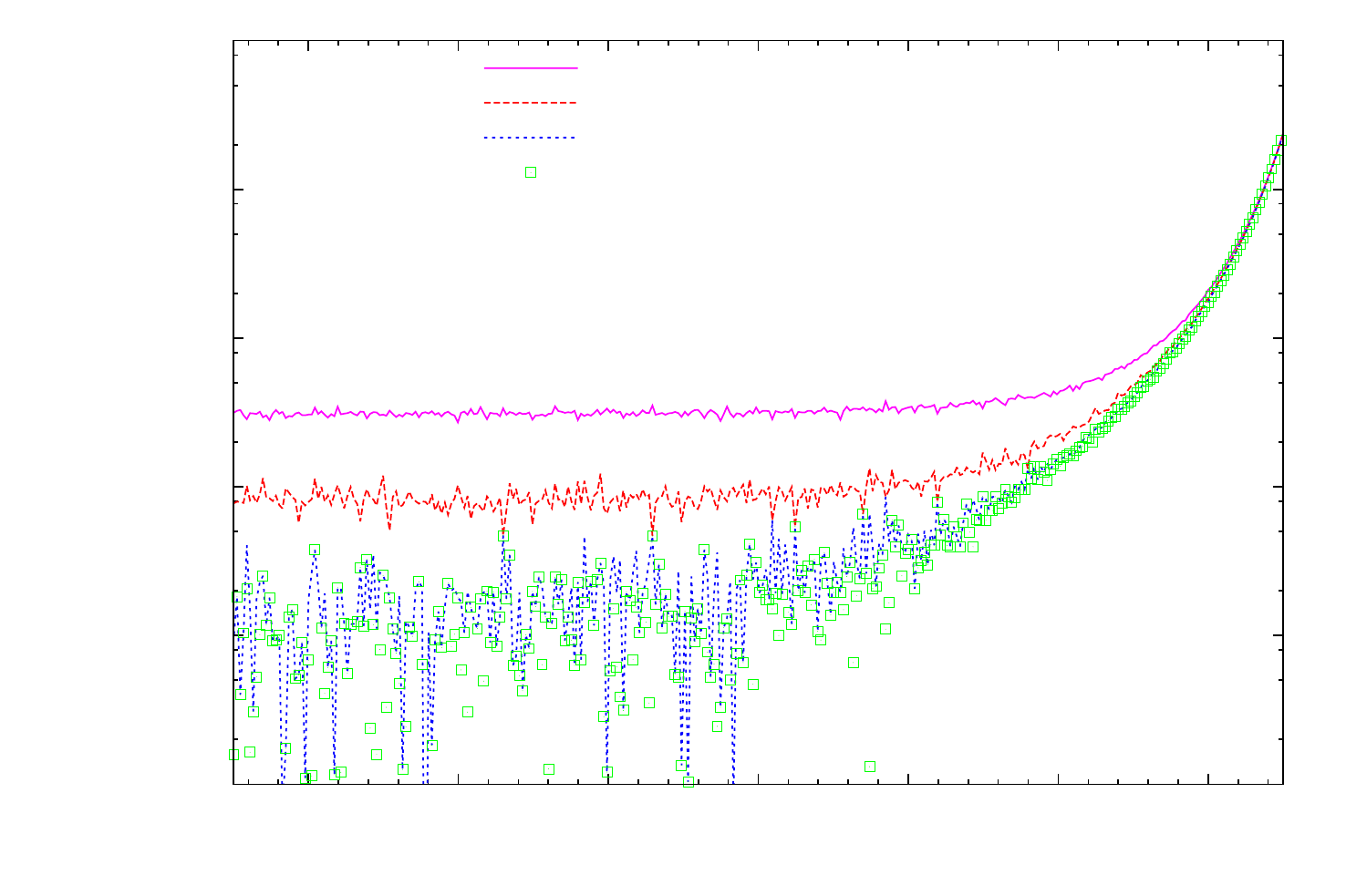}}
\caption{Relative error of {proximity $P(\tau)$} compared to the asymptotic formula (\ref{asym:Tstar}) plotted against $T^*-t$ (bottom axis) and $\tau$ (top axis). Curves show convergence in $d\tau$ and an error $\sim10^{-7}$ persistent to exceptionally small values of $T^*-t$. \label{fig:Tstar1}}
\end{figure}

However, one should proceed with caution when dismissing the ability of the mapped system at assessing its original temporal position: error in the assessment of $T^*$ is not to be confused with error in the \emph{proximity} to $T^*$. Although the original system can integrate to within $10^{-14}$ of $T^*$, it is impossible to assess any behaviour beyond this point: this is a solid barrier for the method. For the mapped system this is not the case: through $t(\tau)$ we can produce an estimate for $T^*-t$ as a function of $\tau$ by considering the following { `proximity'} integral
\[ T^*-t \approx P(\tau)= \int_\tau^\infty \frac{1}{\|\gamma(\cdot,t(\tau'))\|_\infty} \mathrm{d} \tau'.\]
Recalling that $\|\gamma(\cdot,t)\|_\infty $ is recovered from  $\langle\gamma_{\mathrm{map}}^2\rangle$ in the mapped system {via formula (\ref{eq:norm_gamma})}, we fit the behaviour of this global measure in the preceding $(\tau-10)$ window via the ansatz
$ \langle\gamma_{\mathrm{map}}^2\rangle \sim \kappa - \frac{m}{\tau}.$
Inserting this into what is now the double integral for $T^*-t$ we obtain 
\[P(\tau)\approx \frac{e^{\left(1-2\kappa\right)\tau} \tau^{2m}\left(1-2\kappa\right)^{2m}}{\|\gamma(\cdot,\tau)\|_\infty} \Gamma \left(1-2m,(1-2\kappa)\tau \right)\,,\]
where $\Gamma$ is the incomplete gamma function. This provides a running estimate for $T^*-t$ which we can validate against the asymptotic formula, equation (\ref{asym:Tstar}). Figure \ref{fig:Tstar1} shows the relative error in $P(\tau)$ as a function of $T^*-t$ in order to demonstrate how the error depends on the absolute proximity to $T^*.$ We find relative errors of the order of $10^{-7}$ persisting far beyond the machine precision limit, and converging at larger $d\tau$ than $Q_\gamma,$ presumably due to the accuracy of the fitting procedure.

\section{Conclusion and Discussion}
\label{sec:concl}
In this paper we have shown that only by mapping the singular system (\ref{GAMMA2}), (\ref{OMEGA3}) to a regular one (\ref{PDE1}), (\ref{PDE2}), can certain unconventional late-time behaviours be observed and asymptotic trends be established. 
The first unusual feature shown is the slow spatial collapse and unusual (Gaussian) Fourier spectrum very near singularity time. This means that the solution will remain well spectrally converged until extraordinarily close to singularity time for even modest resolutions. In turn this implies a fundamental constraint on the original system: in the original variables one can only hope to approach $T^*$ to the precision of the floating point arithmetic being used, usually double-precision, $\approx10^{-16}.$ Because of this lack of digits in the independent temporal variable, assessing any quantities from the original system is a hazardous undertaking as errors grow exponentially. In other words, not only does the proximity to $T^*$ present a floating point barrier: it also harms the accurate assessment of the late time behaviour of the system before the barrier is reached. On the other hand, the mapped system has no floating point arithmetic barrier as the singularity time is now at infinity and we observe uniform errors until $T^*-t$ is exceptionally small ($10^{-140}$ in the figures shown).\\

Another floating point barrier also becomes apparent, namely that $\| \gamma \|_\infty$ will eventually overflow, i.e. exceed $\sim 10^{308}$ at $\tau\approx 715$. Luckily the mapping allows us to postpone this barrier further by simply computing $\log \| \gamma \|_\infty$ (i.e., outputting the exponent of the right hand side of equation (\ref{eq:norm_gamma}))and use an arbitrary precision exponential in post-processing if required.\\

There is already some evidence that, depending on the type of initial conditions, the full 3D problem has a changing late-time regime where either a depletion of nonlinearity slows vorticity growth \cite{Hou:2006p2135} or the collision of two vortex sheets accelerates the loss of regularity \cite{Bustamante:2012jc}. It is therefore hoped that mapping the full 3D problem will give renewed confidence in the late-time behaviour of the next generation of simulations of 3D Euler.


\begin{acknowledgments}
This publication has emanated from research supported in part under the Programme for Research in Third Level Institutions (PRTLI) Cycle 5; the European Regional Development Fund; and a research grant from Science Foundation Ireland (SFI) under Grant Number 12/IP/1491. Computational resources were provided by the Irish Centre for High-End Computing via class C projects ndmat023c and ndmat025c.
\end{acknowledgments}

\bibliography{papers}
\bibliographystyle{pnas2011}

\end{article}



\end{document}